 \documentclass[12pt,showpacs,preprintnumbers,amsmath,amssymb]{revtex4}
 \usepackage{graphicx}

\begin{document}

\title{Nonlinear effects at the Fermilab Recycler e-cloud instability.}
\author{V.~Balbekov}
\affiliation {Fermi National Accelerator Laboratory\\
P.O. Box 500, Batavia, Illinois 60510}
\email{balbekov@fnal.gov} 
\date{\today}
\begin{abstract}
\begin{center}
{\bf Abstract}
\end{center}

Theoretical analysis of e-cloud instability in the Fermilab Recycler is 
represented in the paper. 
The e-cloud in strong magnetic field is treated as a set of immovable snakes 
each being initiated by some proton bunch.    
It is shown that the instability arises because of injection errors of the bunches 
which increase in time and from bunch to bunch along the batch.
being amplified by the e-cloud electric field. 
The particular attention is given to nonlinear additions to the cloud field.
It is shown that the nonlinearity is the main factor which restricts growth 
of the bunch amplitude.
Possible role of the field free parts of the Recycler id discussed as well.
Results of calculations are compared with experimental data demonstrating 
good correlation. 

\end{abstract}
\maketitle

%
\newpage
\section{Introduction}
%

Fast coherent instability of horizontal betatron oscillations of bunched proton beam 
was observed in the Fermilab Recycler since 2014 as it is described 
in Ref.~\cite{MAIN}.
It has been shown in this paper that the instability is caused by electron 
cloud which arises at ionization of residual gas by protons, 
and grows later due breeding of the electrons at collision with the beam pipe 
walls.

A theoretical model of the instability has been proposed in 
Ref.~\cite{MY}.
The electron cloud is treated as a set of ``snakes'' each of them
appearing as a footprint of some proton bunch. 
The snakes are immovable in horizontal plane due to strong vertical 
magnetic field.
However, the electrons are very mobile in vertical direction because they 
move between the beam pipe walls under the influence of electric field 
of the protons.
They can breed or perish at collisions with the beam pipe walls.

The model provides a suitable description of initial part of the instability
including dependence of the bunch amplitude on time and the position in the batch.
However, it predicts an unrestricted growth of the bunch amplitudes 
which statement is in conflict with the experimental evidence.
It follows from the experiment that the amplitude increases with variable growth rate
within 60-80 turns and becomes about stable after that.

It was suggested in \cite{MY} that nonlinearity of the e-cloud field can 
be responsible for similar behavior of the proton beam, and several examples 
have been represented there. 
The development of this idea is a subject if this paper.
It is shown the it is a way to bring the calculation into accordance with the 
experimental evidence.  

%

\section{Electron cloud model}

%
\begin{figure}[t!]
\includegraphics[width=100mm]{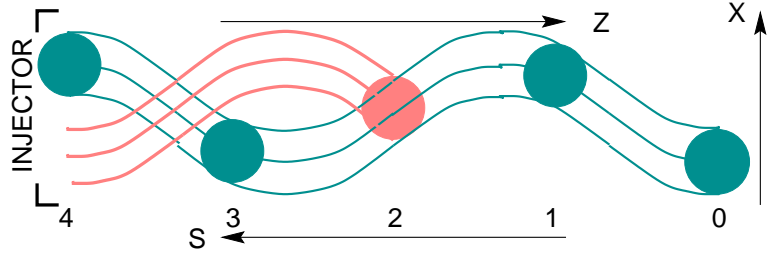}
\caption{Top view of the e-cloud. Each proton bunch gives rise to an 
immovable e-snake.
The snakes coincide with each other if their parent proton bunches 
have the same injection conditions being different otherwise~(\#2 in the picture). 
Local density of each snake depends on time.}
\end{figure}
%

It has been shown in \cite{MY} that horizontal motion of electrons in the 
cloud is awfully obstructed by the Recycler magnetic field.
Vertical motion between the walls and the electron breeding in the walls 
result in creation of a vertical strips \cite{MAIN} 
and in the formation of the ``snake'' as it is shown in Fig.~1.
Each proton bunch creates the wake following the bunch in accordance with 
the injection error.
The bunch wakes coincide if they are injected with the same error
(\#0, 1, 3, 4 in Fig.~1).
Any wake has a steady shape but variable density dependent on time.

According to this model, electron density at distance $\,s\,$ from  
beginning of the batch can be represented in the form
%
\begin{equation}
 \rho_e(x,s,t) = \int_0^s w\left(\frac{s-s'}{v}\right)\,\bar\rho
 \left(x-X\Big(s,t-\frac{s-s'}{v}\Big)\right)\lambda(s')\,ds'
\end{equation}
%
where $\,\bar\rho(x)\,$ is normalized projection of the proton steady state
distribution on axis $\,x$, $\,X(s,t)\,$ is the beam coherent displacement,
and $\,\lambda(s)\,$ is its linear density. 
The coefficient $\,w(\tau)\,$ describes evolution of the snake local density 
which has been considered in Ref.~\cite{FUR}-\cite{PIV}.
Calculation of this function is not a subject of this paper, and it will be
treated further as some phenomenological parameter.

Because the electron distribution is flat in $(y$-$z)$ plane, and effect of the 
walls is small within the proton beam, electric field of this beam is
%
\begin{eqnarray}
 E_e(x,s,t) = e\int_0^s w\left(\frac{s-s'}{v}\right)\,F\left(x-X\Big(s,t
-\frac{s-s'}{v}\Big)\right)\lambda(s')\,ds'
\end{eqnarray}
%
with the function F satisfying the equation  
%
\begin{eqnarray}
 F'(x) = 4\pi \bar\rho(x)
\end{eqnarray}
%
If the beam consists of short identical bunches, the integral turns into the sum
%
\begin{equation}
 E_n(t,x) = eN_b\sum_{m=0}^{n} w_{k} F\Big(x-X_{n-m}(t-mT_{RF})\Big)
\end{equation}
%
where $\,N_b\,$ is the bunch population, $\,T_{RF}\,$ is the time separation 
of the bunches which are enumerated from the beam head (index 0) to the 
current bunch (index $n$).

%

\section{Proton equation of motion}

%
With the cloud electric field taken into account,  
equation of horizontal betatron oscillations of a proton in $\,n^{\rm th}$ bunch is  
%
\begin{equation}
 [\ddot x(t)+\omega_0^2x]_n =
-\frac{e^2N_b}{m\gamma}\sum_{m=0}^{n}w_{m}F\Big(x-X_{n-m}(t-mT_{RF})\Big)
\end{equation}
%
where $\omega_0$ is betatron frequency without e-cloud
(we do not consider here other factors which could affect the betatron 
motion, for example chromaticity).

Because $\,\bar\rho(x)\,$ is the odd function, approximate solution of Eq.~(3) 
including the lowest nonlinearity is
%
\begin{equation}
 F(x)\simeq 4\pi \bar\rho(0)\left(x +\frac{\epsilon x^3}{3}\right),\qquad
 \epsilon = \frac{1}{2\bar\rho(0)}\frac{d^2\rho(0)}{dx^2} 
\end{equation}
%
Therefore equation of betatron oscillations 
of a proton in $\,n^{\rm th}\,$ bunch obtains the form
%
\begin{equation}
 [\ddot x(t)+\omega_0^2x(t)]_n=-2\omega_0\sum_{m=0}^n W_m\xi_m
 \left(1+\frac{\epsilon_m\xi_m^2}{3}\right),
 \qquad\xi_m = x(t)-X_{n-m}(t-nT_{RF})
\end{equation}
%
where $\,W_m=4\pi e^2\bar\rho_m(0)w_m/(m\gamma\omega_0)$. 
Without coherent oscillations that is at $\,X_j=0\,$, equation of small 
incoherent oscillations of protons in $\,n^{\rm th}\,$ bunch is  
%
\begin{equation}
 \ddot x(t)+\omega_n^2x(t)=0, \qquad\omega_n=\omega_0+\Delta Q_n, \qquad
 \Delta Q_n=\sum_{m=0}^nW_m.
\end{equation}
%
It means that $\Delta Q_n$ is the incoherent tune shift of protons in 
$\,n^{\rm th}$ bunch caused by e-cloud produced by all foregoing bunches, 
and $\,W_m\,$ is the contribution of the bunch \#$(n-m)$. 

%

\section{Linear approximation}

%

At $\,\epsilon_m=0$, Eq.~(7) can be averaged over all particles of 
$\,n^{\rm th}\,$ bunch 
resulting series of equations for coherent oscillations of the bunches 
%
\begin{equation}
 \ddot X_n(t)+\omega_0^2X_n=-2\omega_0\sum_{m=0}^{n} 
 W_{m}\Big[X_n(t)-X_{n-m}(t-mT_{RF})\Big]
\end{equation}
%
This series has been investigated in detail in Ref.~\cite{MY}.
The main conclusions of the paper are summarized below and illustrated 
by Fig.~(2).

1. Injection errors are the root cause of the ``instability''. 
The initial amplitude can increase in time as well as from bunch to bunch 
along the batch. 

2. Some spread of the errors is another condition for the instability.
Otherwise solution of Eq.~(9) is $\,X_n(t)=X_0(t-nT_{RF})$
that is all bunches move one by one along the same stable trajectory.  
Coherent interaction of the bunches is absent at such conditions.

3. A variability of the wake is another condition of the instability because the 
bunches have different eigentunes and their resonant interaction is impossible
at $\,W_m=\,$const.

4. With restricted wakes, the eigentunes have the same value in the batch tail
where the amplitude growth should be maximal.
This statement is in agreement with experimental evidence.  

5. Dependence of the amplitude on time is non-exponential generally
being different from bunch to bunch.

6. However, growth of amplitudes is unrestricted at long last, which conclusion 
contradicts the experimental evidence. 
Therefore this statement requires an analysis beyond the scope 
of linear approximation.  

%
\begin{figure*}[b!]
\hspace{-20mm}
\begin{minipage}[h!]{0.45\linewidth}
\begin{center}
\includegraphics[width=85mm]{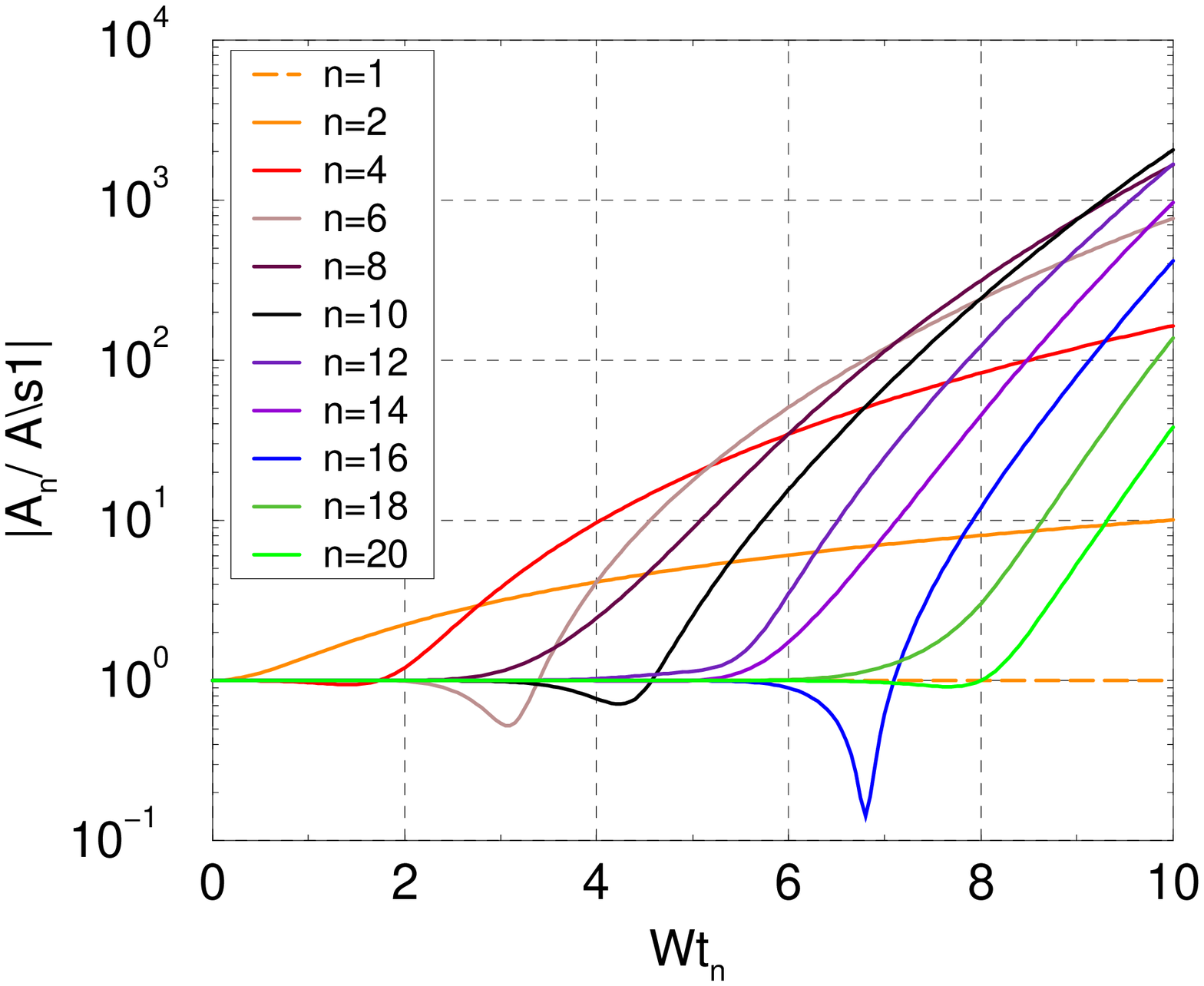}
\end{center}
\end{minipage}
\hspace{-5mm}
\begin{minipage}[h!]{0.45\linewidth}
\begin{center}
\includegraphics[width=85mm]{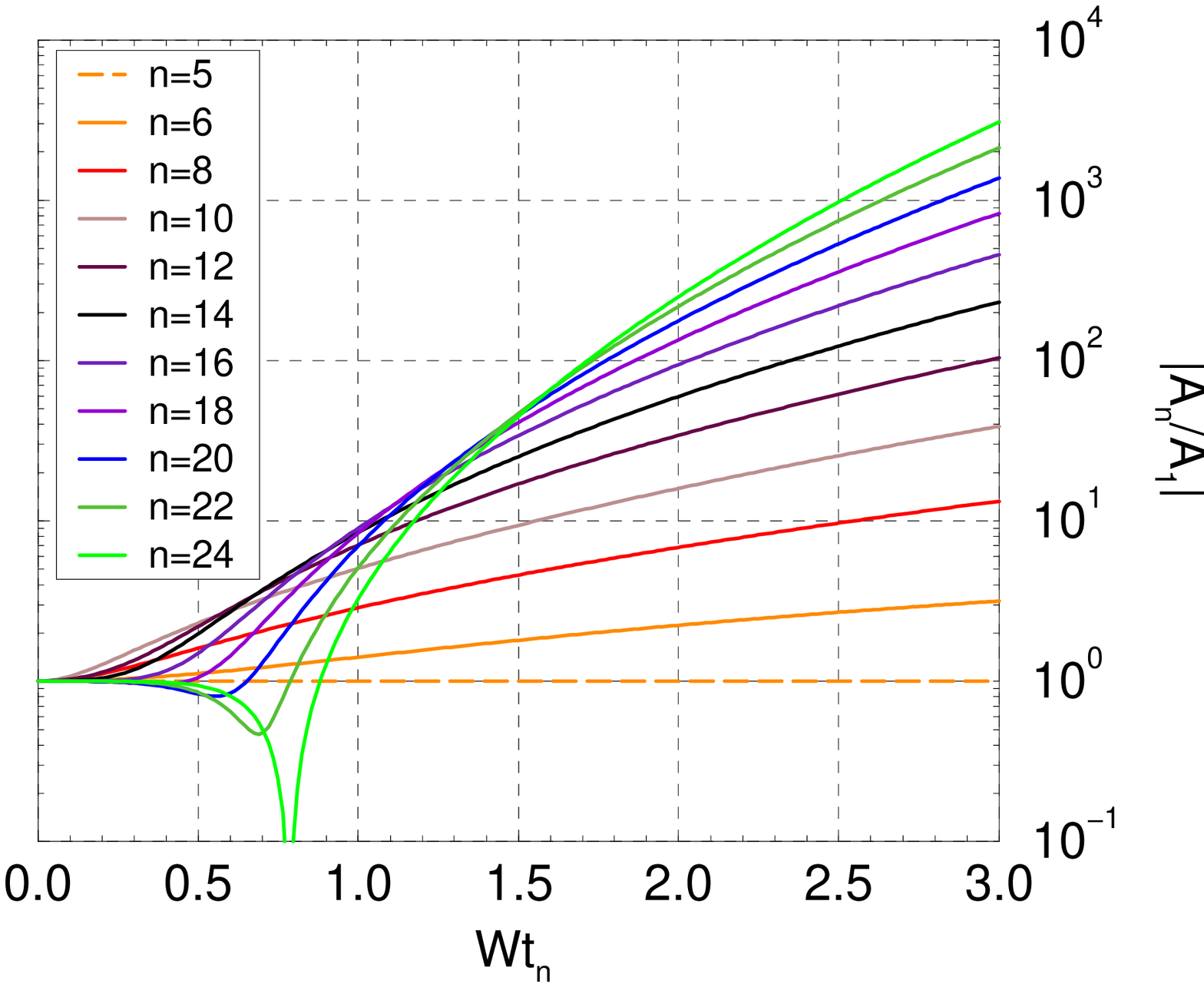}
\end{center}
\end{minipage}
\caption{E-cloud instability in linear approximation. 
Left-hand graph represents effect one-step wake $\,W_n=W\delta_{n,1}$, 
right-hand graph refers to the 5-steps wake.
Index $\,n\,$ is the bunch number in the batch, 
$\,t_n=t-nT_{RF}\,$ with $\,t\,$ as the current time.
It is assumed that the leading bunches are not oscillating: 
$\,A_0=0,\,$ at the left-hand plot, 
and $\,A_{0-4}=0\,$ at the right-hand one.
Initial amplitude of other bunches $\,A_n(0)=1\,$.}
\end{figure*}
%

%

\section{Nonlinear consideration} 

%

We will represent the variables $\,x\,$ and $\,X\,$ in Eq.~7 with help of 
the complex amplitudes $\,a\,$ and $\,A$:
%
\begin{equation}
 x(t) = a(t)\exp\big(i\omega_0[t-nT]\big)+c.c.,\qquad
 X_m(t) = A_m(t)\exp\big(i\omega_0[t-mT]\big)+c.c.
\end{equation}
%
Substituting these values in Eq.~(7) and applying the standard method of averaging, 
one can get following equations for amplitude of a proton inside 
$\,n^{\rm th}$ bunch \cite{MY}
%
\begin{equation}
 \dot a(t) = i\sum_{m=0}^n W_m\eta (1+\epsilon_m|\eta_m|^2),    
 \qquad \eta_m=a(t)-A_{n-m}(t-mT).
\end{equation}
%
One-step wake will be investigated further: $\,W_n=W\delta_{n1}$.
Note that the condition $\,W_0=0$ follows from this definition 
being very reasonable because a noticeable e-cloud cannot appear in 
the leading bunch without secondary electrons.
Therefore any proton has a constant betatron amplitude in this bunch, 
and the same is valid for the bunch coherent amplitude as well.
The last can be taken as $\,A_0=0\,$ because difference of the bunch amplitudes 
is the only crucial circumstance.
With these approximations, equations of motion of any proton inside 
$\,n^{\rm th}$ bunch is 
%
\begin{equation}
 \dot a(t) = iW\big[\,a(t)-A_{n-1}(t-T)\,\big]
 \big[\,1+\epsilon\big|a(t)-A_{n-1}(t-T)|^2\big],
 \qquad A_0=0.
\end{equation}
%
Following steps have to be used for numerical solution of these equations:

1. To generate a random initial distribution of particles in first bunch $(N=1)$.
   The bunch central amplitude should be $\,A_1(0)\ne 0\,$ to begin the process.

2. To calculate the function $\,a(t)\,$ for each particle of the first bunch 
   $(n=1)$ by solution of Eq.~(12) with the known value of the amplitude 
   $A_{n-1}=A_0=0$.

3. To calculate the central amplitude $\,A_1(t)\,$ as a function of time
   by the averaging over all particles of the bunch;  

4. To repeat the operation for second bunch with known $\,A_1(t)$, etc.
\\
Results of the calculation are represented below.

%
\newpage
\subsection{Physics of the phenomenon}

%
The linear approximation for the one-step wake has been commented 
in Sec.~IV being represented by left-hand Fig.~2.
At present the same case will be investigated with nonlinear additions
taken into account.
 
Initial amplitude of all bunches except as the leading one $\,A_{n\ne0}(0)=1$, 
and the nonlinear parameter given by Eq.~(6) is taken as large as 
$\epsilon=-0.001$.
The proton beam is considered as thin one that is its radius is assumed 
to be small in comparison with the injection errors.

Obtained coherent amplitude of the bunches is represented in Fig.~3
against the normalized time.
In the beginning, it is about the same as it has been shown in Fig.~2.   
However, further behavior is strongly different.
It is seen that the growth of the bunch amplitudes ceases at about 
$\,|A_n/A_1|=20$ - 30~~which limit is achieved at $\,Wt=8$ - 10. 
%
\begin{figure*}[b!]
\includegraphics[width=85mm]{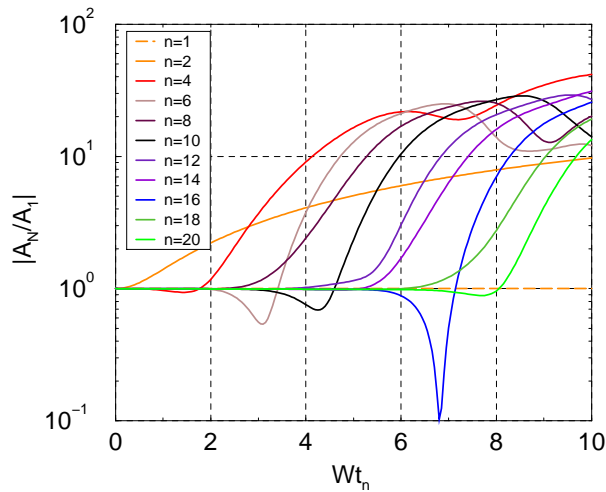}
\caption{Instability with nonlinear e-cloud field. 
The same conditions as in the left-hand Fig.~2 
but nonlinear Eq.~12 is used with the nonlinear parameter 
$\,\epsilon A_1^2=-0.001$.}
\end{figure*}
%

The saturation cannot be treated as Landau damping because thin proton
beam with negligible incoherent tune spread could not be an object of this
phenomenon.
Therefore the nonlinearity does not prevent the instability in the case, 
but merely restricts its growth. 
This statement is illustrated by Fig.~4 where behavior of second bunch of the 
batch is considered in more details. 
The leading bunch does not oscillate as it was assumed, and the first bunch has 
constant amplitude because there is no external force to excite it.
The relative amplitude of second bunch is shown in the left-hand graph against time
at different nonlinearity, and several phase trajectories are represented in the 
right-hand figure.
It is a typical behavior of nonlinear oscillator exited by periodical external
field. 
%
\begin{figure*}[h!]
\hspace{-10mm}
\begin{minipage}[h!]{0.45\linewidth}
\begin{center}
\includegraphics[width=85mm]{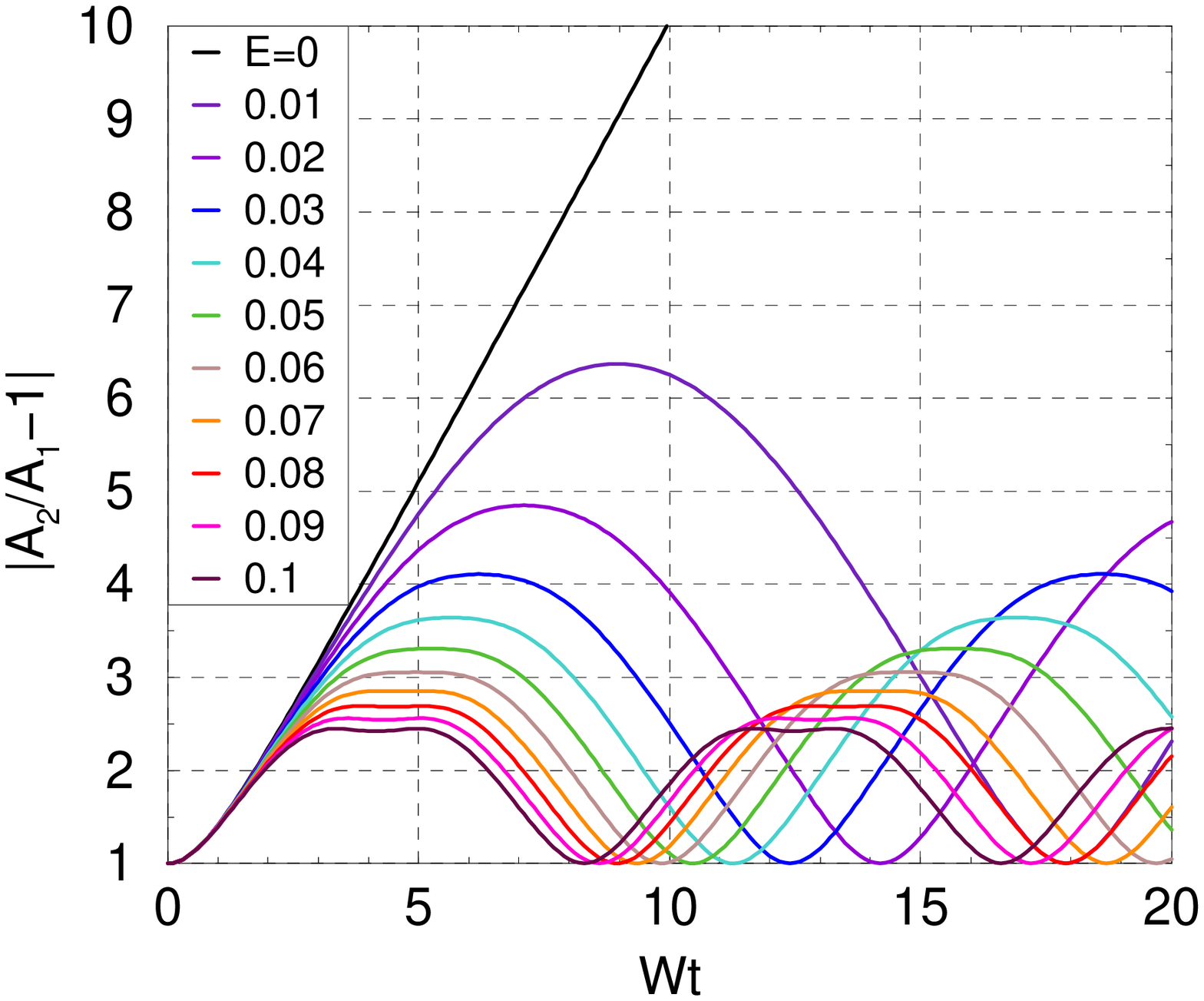}
\end{center}
\end{minipage}
\hspace{0mm}
\begin{minipage}[h!]{0.45\linewidth}
\begin{center}
\includegraphics[width=85mm]{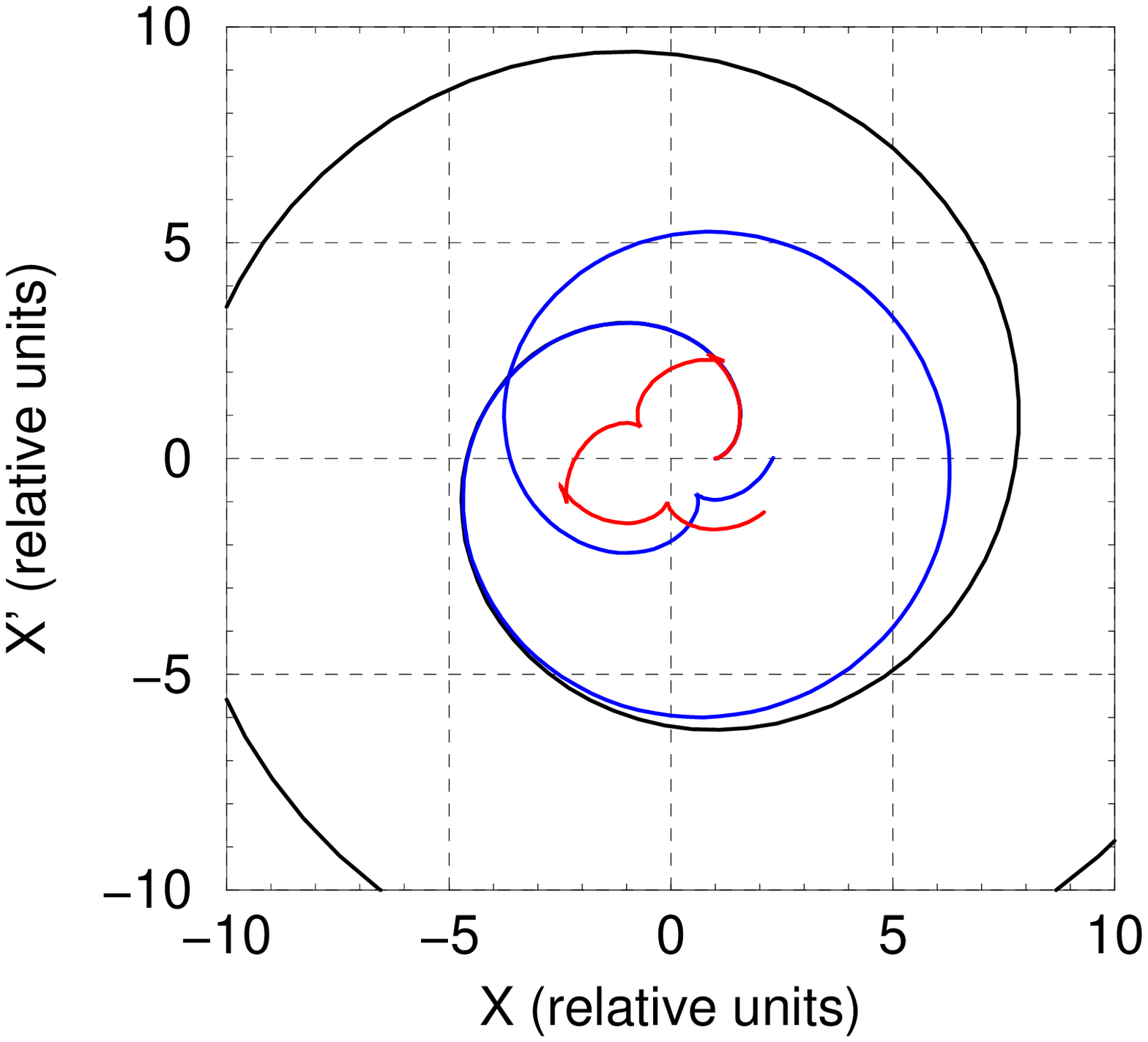}
\end{center}
\end{minipage}
\caption{Second bunch in the train. 
The leading bunch does not oscillate, the first bunch has constant amplitude, 
and the second one has the same initial amplitude: $\,A_2(0)=A_1$.
Its relative amplitude is shown in the left-hand graph against time
at different nonlinearity, and several phase trajectories are presented in the 
right-hand figure.}
\vspace{-5mm}
\end{figure*}
%


\subsection{Dependence on value of the nonlinearity}


Two more examples are represented in Fig.~5. 
In the case, the batch has the same arrangement and initial conditions as in 
Fig.~3 but other parameters of the nonlinearity: 
$\,\epsilon=10^{-4}\,$ and $\,10^{-2}$.
As one can expected, the more nonlinearity results in the less coherent amplitude.
The ultimate amplitude can be estimated by the relation $\,\epsilon A^2\simeq-1$,
and it is attained at about $\,Wt=8-10$.
%
\begin{figure*}[h!]
\hspace{-20mm}
\begin{minipage}[h!]{0.45\linewidth}
\begin{center}
\includegraphics[width=85mm]{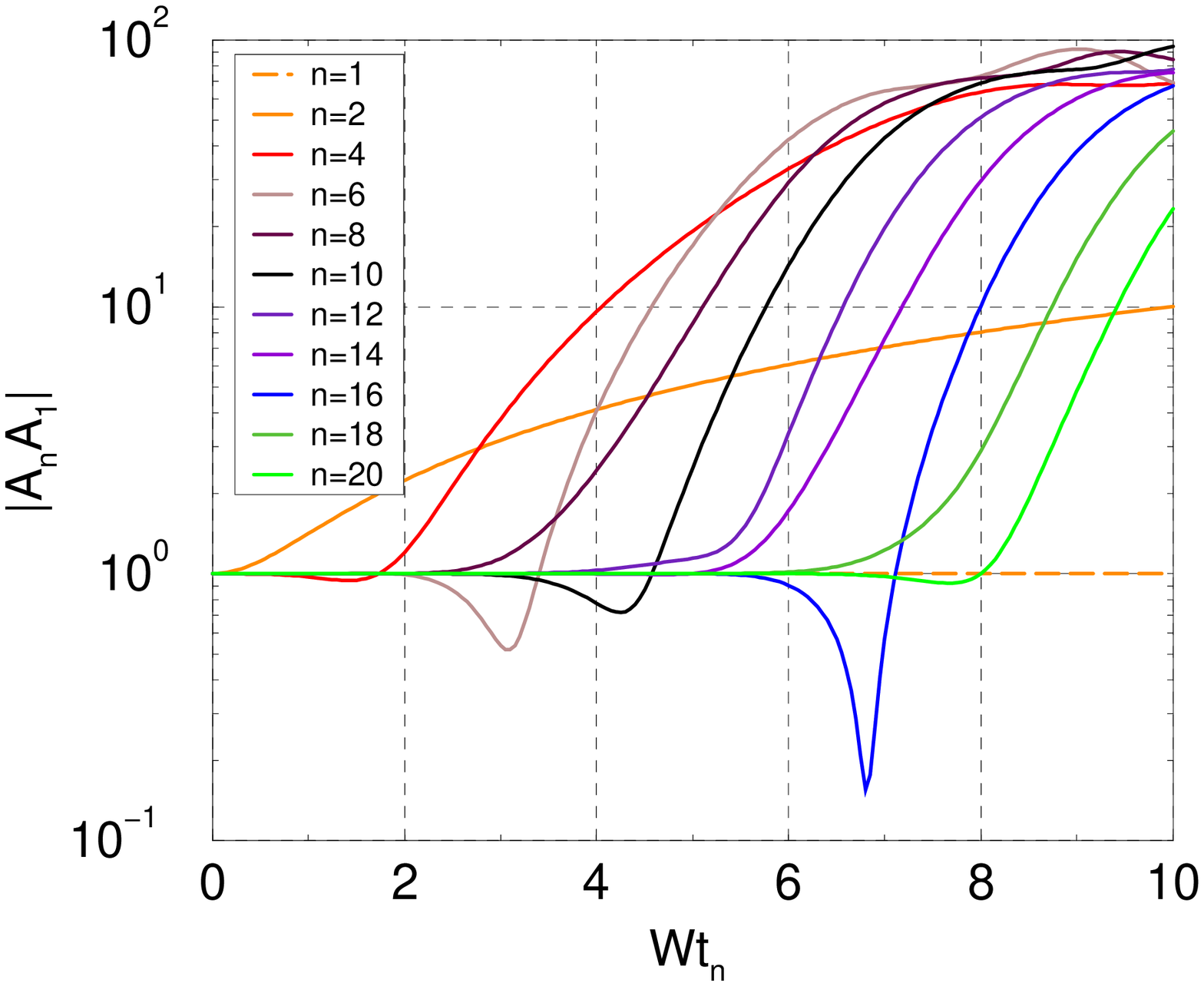}
\end{center}
\end{minipage}
\hspace{-5mm}
\begin{minipage}[h!]{0.45\linewidth}
\begin{center}
\includegraphics[width=85mm]{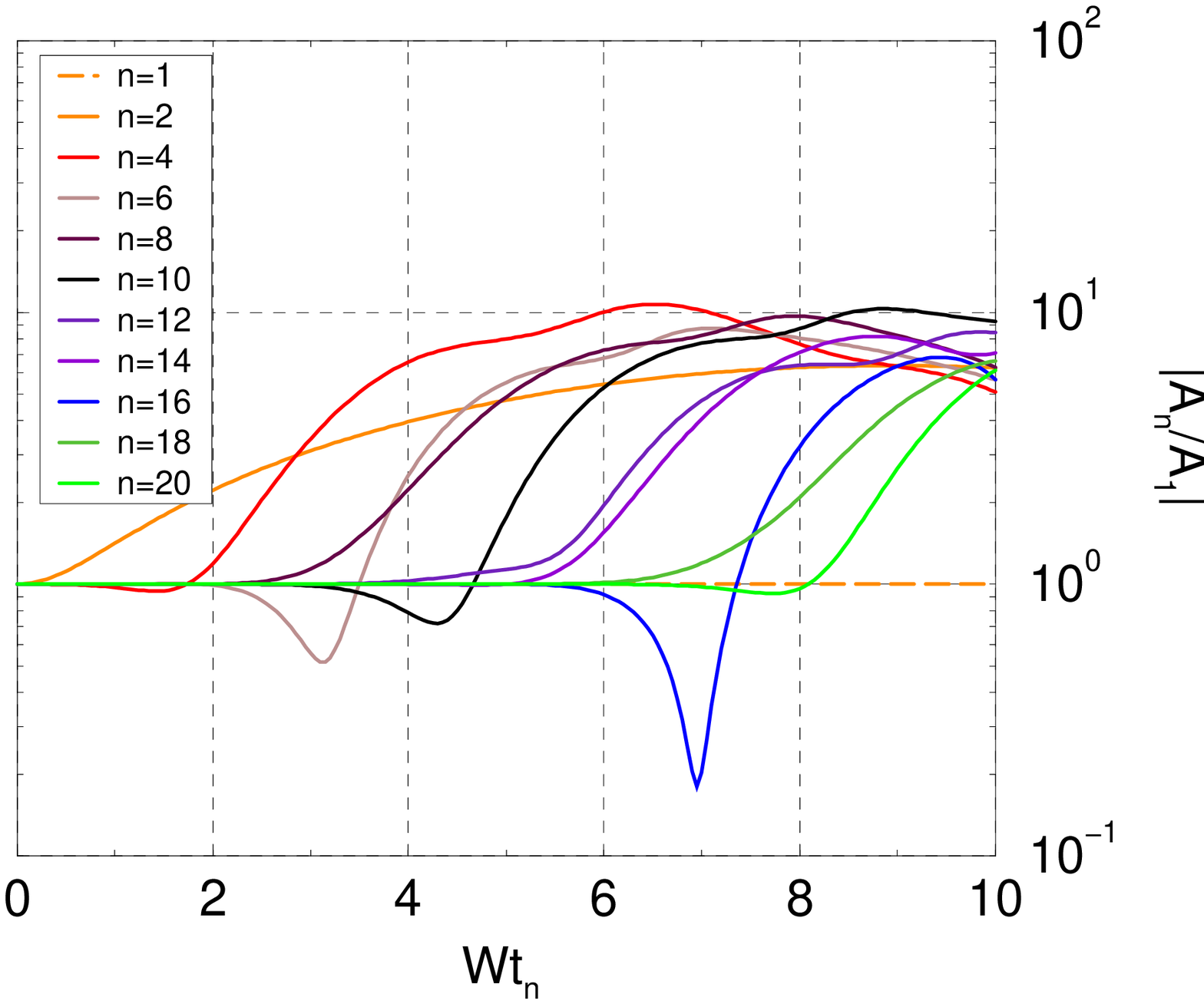}
\end{center}
\end{minipage}
\caption{Instability with nonlinear e-cloud field. 
The same conditions as in Fig.~3 with other nonlinearity:
left-hand graph: $\,|\epsilon A_1^2|=10^{-4}$,
right-hand one:  $\,|\epsilon A_1^2|=10^{-2}$.}         
\end{figure*}
%
\begin{figure*}[h!]
\includegraphics[width=85mm]{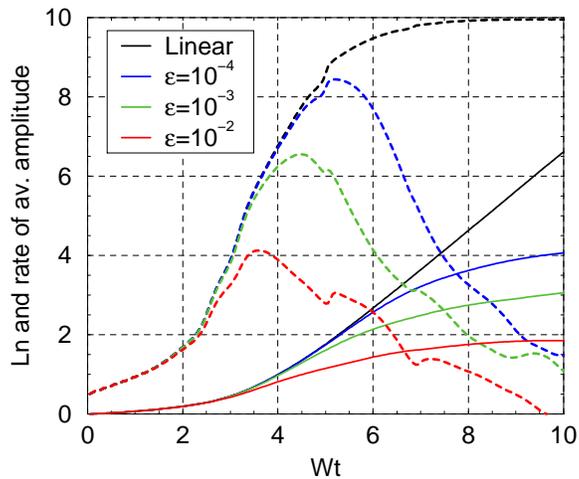}
\caption{The betatron amplitude (solid lines) and its growth rate (dashed lines)
averaged across the batch at different nonlinearity.
Fig.~2, 3, and 5 are used as the sources.}
\end{figure*}
%

The results are summarized in Fig.~6 where averaged across the batch parameters 
are shown.
Solid lines represent the averaged coherent amplitude, and dashed lines -- 
its instantaneous rate
(it is just a picture which has been measured in the experiment \cite{MAIN},
and corresponding comparison will be made later).
Four cases are considered in this example being taken from Fig.~2 (left), 3, 
and 5.  
It is seen that the amplitude growth has about exponential behavior
only at zero nonlinearity, and only at $\,Wt>\sim 5$. 
The nonlinearity does not reveal itself at $\,Wt<\sim 3\,$ 
but restricts the amplitude growth at $\,Wt>\sim 6$ - 10.
The maximal growth rate is about $\,\sim 1/\ln|\epsilon| A_1^2$,
and the maximal amplitude is  $\,A_{max}^2 \sim 0.5/|\epsilon|$.   

%
\newpage
\subsection{Dependence on the beam radius}

%
\begin{figure*}[t!]
\hspace{-20mm}
\begin{minipage}[h!]{0.45\linewidth}
\begin{center}
\includegraphics[width=85mm]{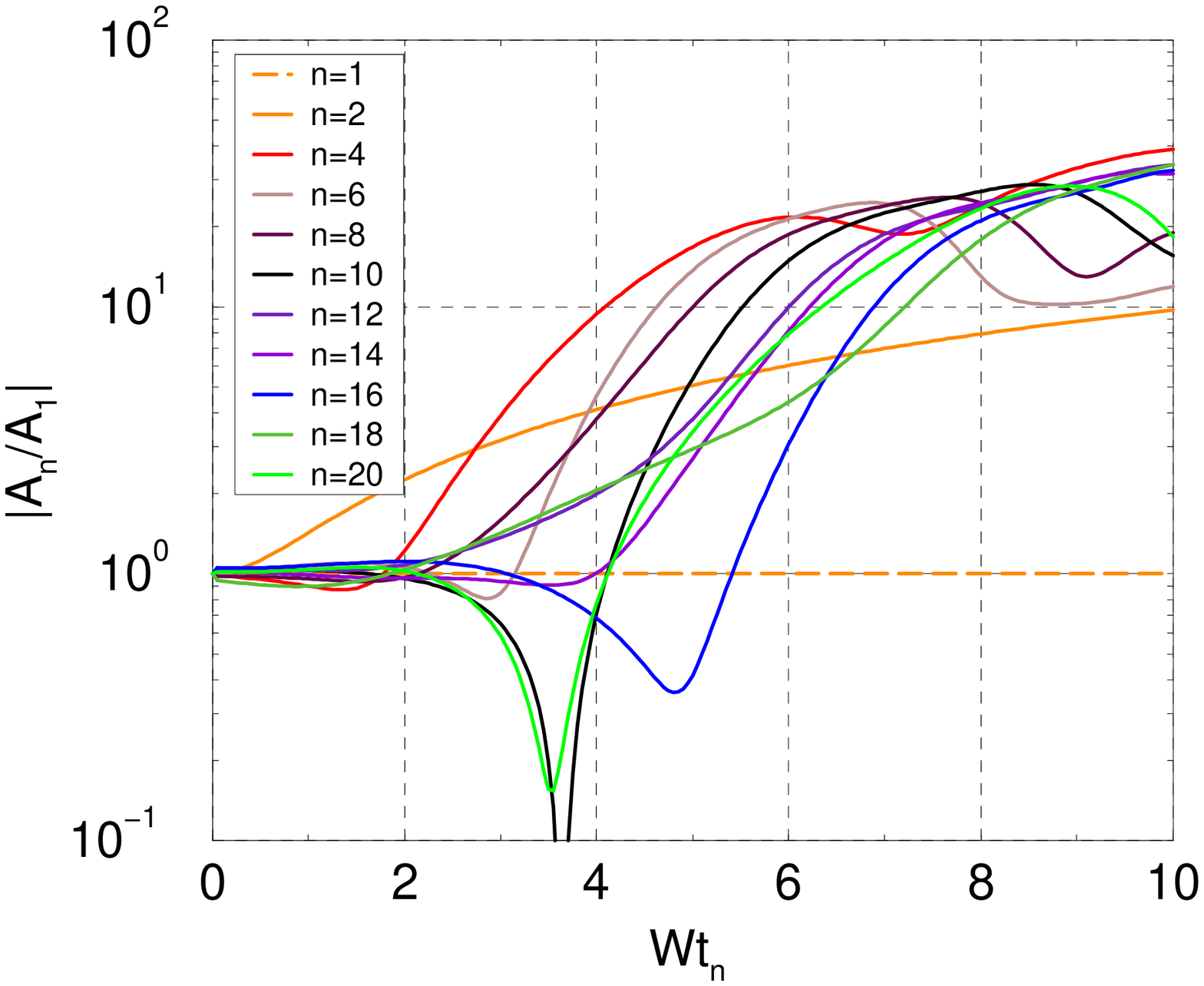}
\end{center}
\end{minipage}
\hspace{-5mm}
\begin{minipage}[h!]{0.45\linewidth}
\begin{center}
\includegraphics[width=85mm]{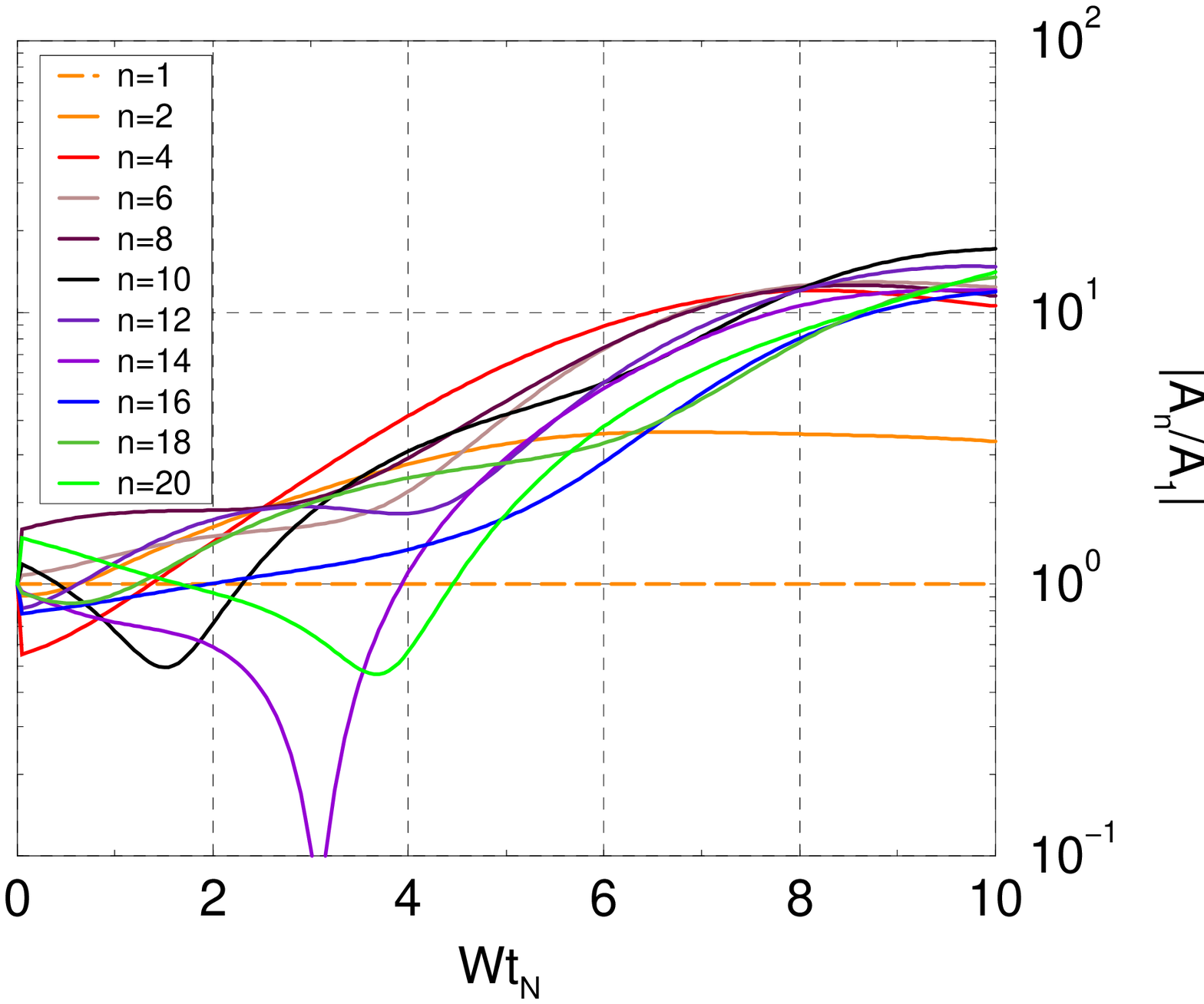}
\end{center}
\end{minipage}
\caption{Instability of thick beam at e-cloud nonlinearity $\,\epsilon=-0.001\,$.   
The same conditions as in Fig.~3 are used with the proton beam radius
$\,R = A_1\,$ in the left-hand graph and $\,R=10A_1\,$ in the right-hand one.}
\end{figure*}
%
\begin{figure*}[h!]
\includegraphics[width=85mm]{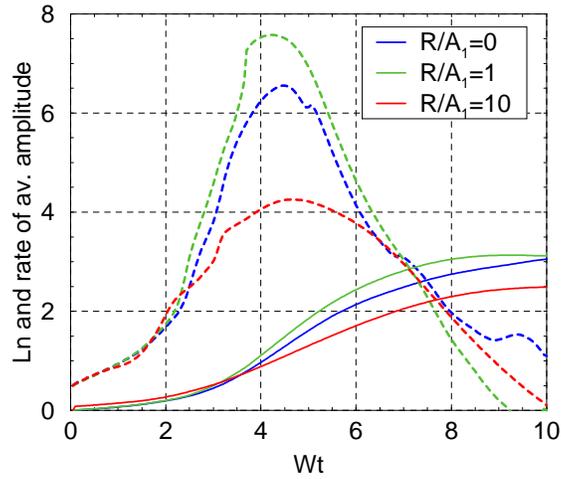}
\caption{The betatron amplitude (solid lines) and its growth rate (dashed lines)
averaged across the batch at different beam radius.
Fig.~3 and 7 are used as the sources.}
\end{figure*}
%

Thick beam is considered in this subsection at the same conditions 
as it has been done in previous part.
The water-bag model of radius $\,R\,$ is used for transverse distribution of the 
proton beam.
The injection error is taken to be unity, and parameter of nonlinearity 
$\,\epsilon=-0.001\,$ in all the cases.
The results are represented in Fig.~7 at $\,R/A_1=1\,$ and 10.
Corresponding averaged values are shown in Fig.~8 where the case $\,R=0,$ is 
added being taken from Fig.~3. 
Comparison of these figures with Fig.~6 and 7 allows to conclude that   
the beam radius is a factor of second importance for the problem.

%

\section{Influence of the field free areas}

%
About a half of the Recycler perimeter is occupied by the field free regions 
where the dipole magnetic field is absent.
The electron production and breeding take place in these regions as well as
in the field filled regions.
Therefore, there is no reasons to think that e-cloud density in the
field-free zones essentially differs from the density in the magnetic zones.  
However, there is no an effective mechanism in the free zones to correlate e-cloud 
position with proton beam so firmly as it makes strong dipole magnetic field.
Therefore direct contribution of the field-free zones to the instability 
is expected to be relatively small. 
However, this part can affect the incoherent motion of protons including 
linear and nonlinear tune shift.
The last is especially important because one cannot to exclude an additional 
restriction of the coherent amplitude due to this addition.  

Because this part of the cloud does not follow the proton beam, its distribution 
should depend on $\,x\,$ but not on $\,X$, in the used terminology.
Taking it into account, one can write correspondingly modified Eq.~(7) 
in the form
%
\begin{equation}
 [\ddot x(t)+\omega_0^2x(t)]_n=-2\omega_0 W 
 \left(\xi+\frac{\epsilon_B\xi^3}{3}+\frac{\epsilon_F x^3}{3}\right)
\end{equation}
%
where $\,\xi = x(t)-X_{n-1}(t-T_{\rm RF})$.
This equation describes betatron oscillations of arbitrary proton 
in $\,n^{\rm th}\,$ bunch.
A one-step wake is considered here, and only cubic nonlinearity is 
taken into account (incoherent linear contribution can be included to $\,\omega_0$).
The coefficients $\,\epsilon_B\,$ and $\,\epsilon_F\,$ describe the nonlinearity 
of the field filled (B) and the field free (F) parts with their relative length 
being taken into account. 

%
\begin{figure*}[t!]
\hspace{-10mm}
\begin{minipage}[h!]{0.45\linewidth}
\begin{center}
\includegraphics[width=85mm]{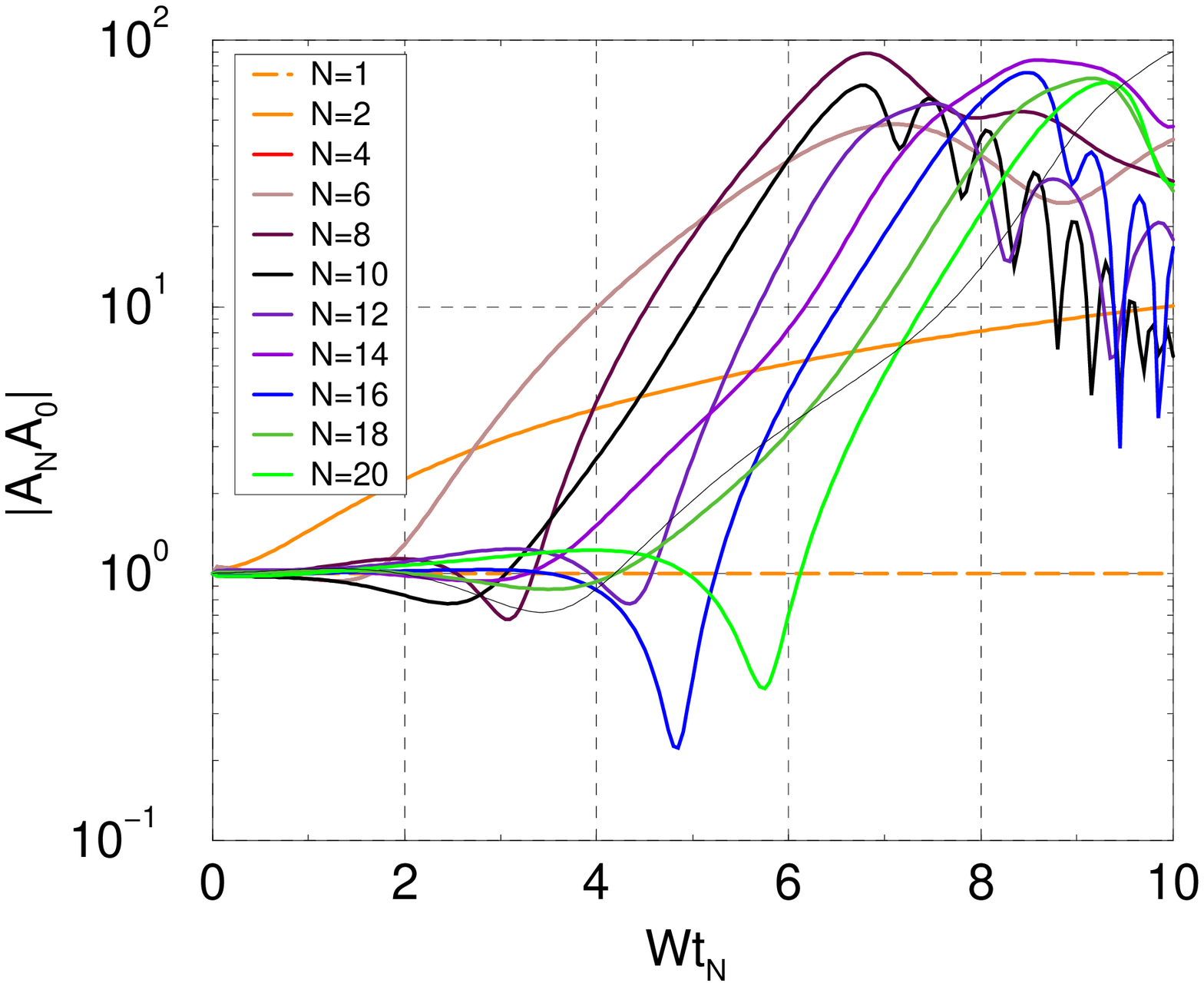}
\end{center}
\end{minipage}
\hspace{0mm}
\begin{minipage}[h!]{0.45\linewidth}
\begin{center}
\includegraphics[width=85mm]{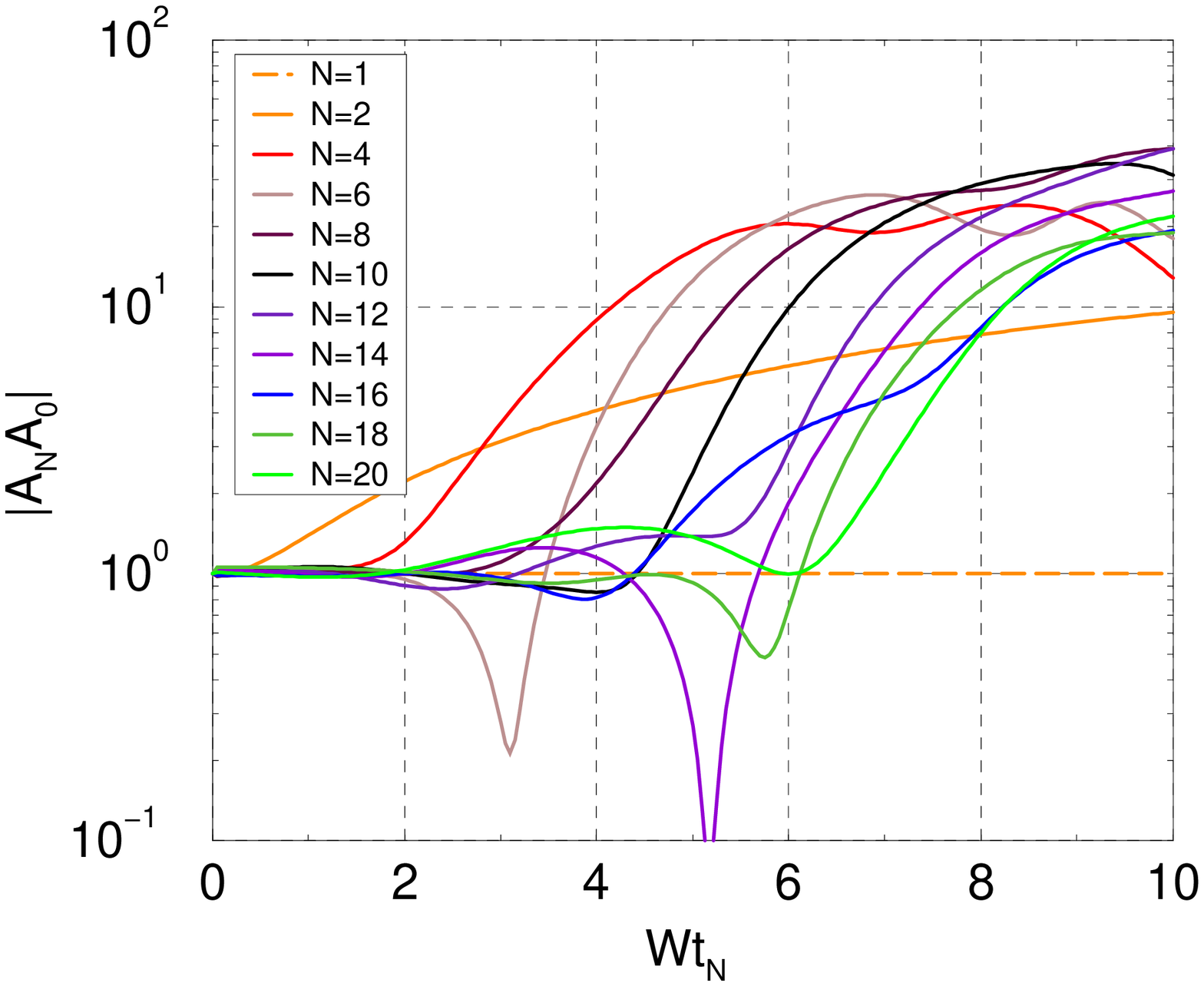}
\end{center}
\end{minipage}
\caption{Instability with field-free regions taken into account.
Contributions of the field filled parts and field free ones are marked 
by symbols B and F. Proton beam radius $\,R=1$, first bunch oscillates 
with amplitude $\,A_1=1$.
Left: $\,\epsilon_B A_1^2=0,\;\epsilon_F A_1^2=-0.001$, right: 
$\,\epsilon_B A_1^2=0=\epsilon_F A_1^2=-0.001$. }
\end{figure*}
%
\begin{figure*}[h!]
\vspace{10mm}
\includegraphics[width=85mm]{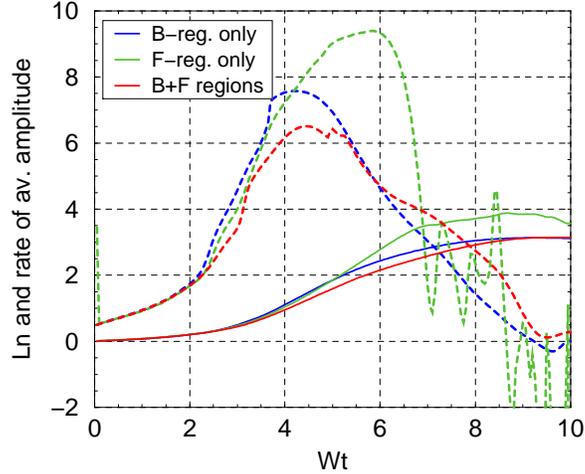}
\caption{The betatron amplitude (solid lines) and its growth rate (dashed lines)
averaged across the batch at different partial nonlinearities.
The beam radius $\,R=1$.
Red: $\,\epsilon_B A_1^2=-0.001,\;\epsilon_F A_1^2= 0$.
Green: $\,\epsilon_B A_1^2= 0,\;\epsilon_F A_1^2=-0.001$.
Blue: $\,\epsilon_B A_1^2=0=\epsilon_F A_1^2=-0.001$. 
Fig.~3 and 9 are used as the sources.}
\end{figure*}
%

Results of the calculations are represented in Fig.~9.
The used beam parameters are: beam radius is taken to be unity, 
leading bunch does not oscillate, injection error of other bunches 
$\,A_n(0)=1\;(n\ne 0)$.

The left-hand Fig.~9 represents the contribution of the field free parts only:
$\,\epsilon_B=0,\;\epsilon_F=-0.001$.
It should be compared with left Fig.~7 where the contribution of the field filled
part has been shown at the same nonlinear parameter. 
It is seen that nonlinearity of the field free parts have less influence on the 
proton coherent oscillations.
It is confirmed by the right-hand Fig.~9 where equal nonlinearities of both kinds 
are considered: $\,\epsilon_B=\epsilon_F=-0.001$. 
It considerably differs from the left-hand figure being rather similar to left 
Fig.~7.
The same conclusion follows from Fig.~10 where the averaged 
beam parameters ate plotted like Fig.~6 and 8.
It is seen that the addition of the field free regions only slightly 
change the results (the blue and the red lines).  

%

\section{Comparison with the experiment}

%
%
\begin{figure*}[b!]
\includegraphics[width=85mm]{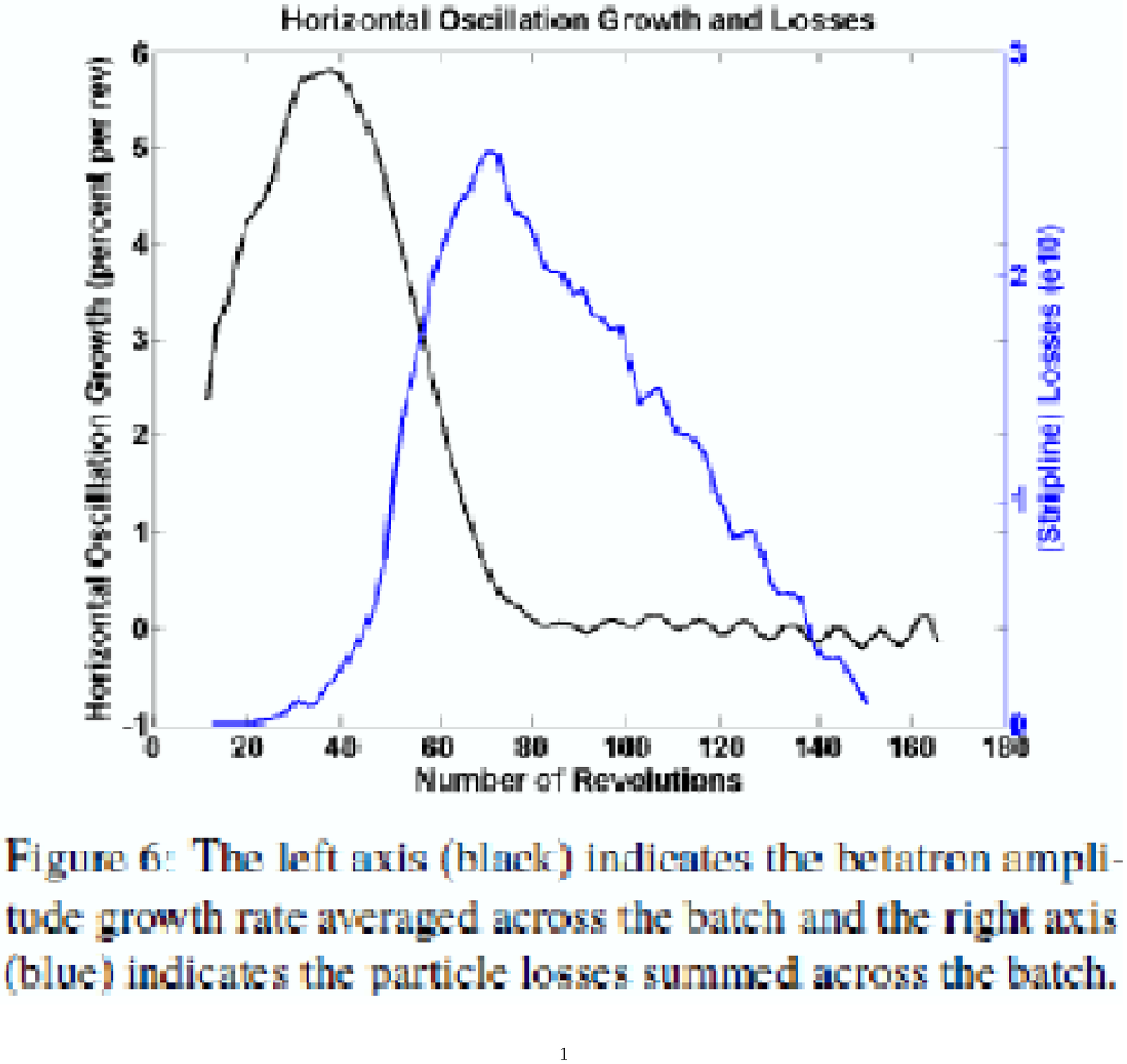}
\caption{}
\end{figure*}
%

Presented results are in a reasonably good agreement with the experimental 
evidence represented in Ref.~\cite{MAIN}.
One of the resumptive plots of this paper is copied and shown here as Fig.11
The black curve in this plot has the same sense as dashed lines in 
Fig.~6 and 8.
All of them demonstrate the instability rate dependent of the parameters which 
can be treated as time measured in different units.
The curves are similar in shape, and the quantitative agreement
can be obtained at following relation of the parameters: 
$$
Wt=10\quad\mbox{corresponds to 80 revolutions, that is}\quad\,WT_{\rm rev}\simeq 1/8
$$
with $\,T_{\rm rev}\,$ as the Recycler revolution time.
On the other hand, it has been shown in Sec.~III that $W$ should be treated 
as betatron tune shift of protons produced by the electron cloud.
It means that  
$$
  WT_{\rm rev}=2\pi\Delta Q\qquad\mbox{that is}\qquad 
  \Delta Q\simeq\frac{1}{16\pi}\simeq 0.02
$$

This result can be used to estimate the central density of the e-cloud $\,n_e$. 
At the accepted model of the cloud, the relation is
%
\begin{equation}
 \Delta Q=\frac{r_0 n_p P^2}{2\pi Q\beta^2\gamma}
\end{equation}
%
where  $\,n_p=1.54\times 10^{-18}$m is the electromagnetic proton radius,
$\,Q=25.45\,$ is the Recycler tune, $\,P=3319\,$m is its perimeter, 
$\,\beta\simeq 1\,$,
and $\,\gamma=9.53\,$ is the normalized energy of protons.
It gives numerically
$$
 \Delta Q \simeq \frac{n_e}{10^{14}{\rm m}^3}\quad
 \mbox {that is}\quad n_e\simeq 2\times 10^{12}{\rm m}^{-3}\quad
 \mbox{at}\quad \Delta Q=0.02
$$
Measurement of the density was not performed in the experiment
but simulation with code POSINS is presented in \cite{MAIN}
resulting in 5-10 times more density.

%

\section{Conclusion}

%

The model of electron cloud in the form of a motionless snake 
is considered in the paper.
Ionization of residual gas by protons is the primary source of the electrons 
being supported by their multiplication in the beam pipe walls. 
Fixation of the electron horizontal position is realized by strong vertical 
magnetic field.
The model allows to explain the electron instability of bunched proton beam 
in the Fermilab Recycler.
According it, the instability is caused by injection errors which initiate 
coherent betatron oscillations of the bunches, 
and electric field of the electron snake promotes an increase of their 
amplitude in time, as well as from the batch head to its tail.
Nonlinearity of the e-cloud electric field is considered in detail
as the important factor restricting the amplitude growth.
The parts of the Recycler perimeter without dipole magnetic field are 
included in the investigation as well.
However, it turns out that their contribution in the instability is negligible.
Results of calculations are in reasonable agreement with the Recycler 
experiment evidence. 


\end{document}